\documentclass[twocolumn,english,aps,prl,reprint,superscriptaddress,showpacs]{revtex4-1}
\usepackage[T1]{fontenc}
\usepackage[latin9]{inputenc}
\usepackage{amstext}
\usepackage{amssymb}
\usepackage{graphicx}

\usepackage{color}



\usepackage{babel}
\begin{document}

\title{Spatiotemporal chaos induces extreme events in an extended microcavity
laser}

\author{F. Selmi}

\affiliation{Laboratoire de Photonique et de Nanostructures, LPN-CNRS UPR20, Route
de Nozay, 91460 Marcoussis, France}

\author{S. Coulibaly}

\affiliation{Laboratoire de Physique des Lasers, Atomes et Molécules, CNRS-UMR8523,
Université de Lille 1, 59655 Villeneuve d'Ascq Cedex, France.}

\author{Z. Loghmari}

\affiliation{Laboratoire de Photonique et de Nanostructures, LPN-CNRS UPR20, Route
de Nozay, 91460 Marcoussis, France}

\author{I. Sagnes}

\affiliation{Laboratoire de Photonique et de Nanostructures, LPN-CNRS UPR20, Route
de Nozay, 91460 Marcoussis, France}

\author{G. Beaudoin}

\affiliation{Laboratoire de Photonique et de Nanostructures, LPN-CNRS UPR20, Route
de Nozay, 91460 Marcoussis, France}

\author{M.G. Clerc}

\affiliation{Departamento de Física, Facultad de ciencias Físicas y Matemáticas,
Universidad de Chile, Casilla 487-3, Santiago, Chile.}

\author{S. Barbay}

\affiliation{Laboratoire de Photonique et de Nanostructures, LPN-CNRS UPR20, Route
de Nozay, 91460 Marcoussis, France}

\email{sylvain.barbay@lpn.cnrs.fr}

\selectlanguage{english}%
\begin{abstract}
Extreme events such as rogue wave in optics and fluids are often
associated with the merging dynamics of coherent structures. We present
experimental and numerical results on the physics of extreme events
appearance in a spatially extended semiconductor microcavity laser
with intracavity saturable absorber. This system can display deterministic
irregular dynamics only thanks to spatial coupling through diffraction
of light. We have identified parameter regions where extreme events
are encountered and established the origin of this dynamics in the
emergence of {deterministic} spatiotemporal chaos, through the correspondence between
the proportion of extreme events and the dimension of the strange
attractor.
\end{abstract}

\pacs{05.45.-a, 42.55.Sa, 42.65.Sf}

\maketitle
A record spawned by a natural system may consist of periods where
a relevant variable undergoes small variations around a well-defined
level provided by its long-time average, with the occasional occurrence
of abrupt excursions to values that differ significantly from the
average level, called extreme events \cite{NicolisBook2012}. Extreme
and rare events are ubiquitous in nature. In optics, an extreme event
is characterized by a rare, intense optical pulse in a given intensity
probability density distribution. The study of extreme events and
extreme waves \cite{OnoratoPR13} has been motivated by the analogy
with rogue waves in hydrodynamics \cite{KharifEJM03} that are giant
waves recently observed in the ocean and whose formation mechanism
is still not well understood. Physically, it is based on the fact
that some conservative systems in optics and deep water waves in ocean
can be described by the nonlinear Schrödinger equation \cite{SolliNature07}.
Most of the studies in this context have taken place in optical fibers
where the interplay of nonlinearity, dispersion and noise generates
extreme events \cite{DudleyOE08,MussotOE09,KiblerNatPhys10,ArecchiPRL11}.
{Extreme events such as rogue wave in optics and fluids are often
associated with the merging dynamics of coherent structures \cite{LecaplainPRL12,AntikainenNL12,BirkholzPRL13},
with stochastically induced transition in multistable systems \cite{PisarchikPRL11} or 
with chaotic dynamics in low dimensional systems \cite{BonattoPRL11}.
Extreme events have been observed} in optical cavity systems,
such as an injected nonlinear optical cavity \cite{MontinaPRL09},
fiber lasers \cite{RandouxOL12,LecaplainPRL12}, solid-state lasers
\cite{KovalskyOL11} and semiconductor lasers \cite{BonattoPRL11,KarsaklianDalBoscoOL13}.
The role of spatial coupling has not been studied until recently in
a pattern forming optical system composed of a photorefractive crystal
subjected to optical feedback \cite{OdentNatHazard10,MarsalOL14}
or low Fresnel number solide-state laser\cite{BonazzolaJO13}, while
most of the characterizations of extreme events were done from a statistical
point of view, without establishing their origin from the dynamical
systems point of view.

\begin{figure}[tbh]
\begin{centering}\includegraphics[width=1\columnwidth]{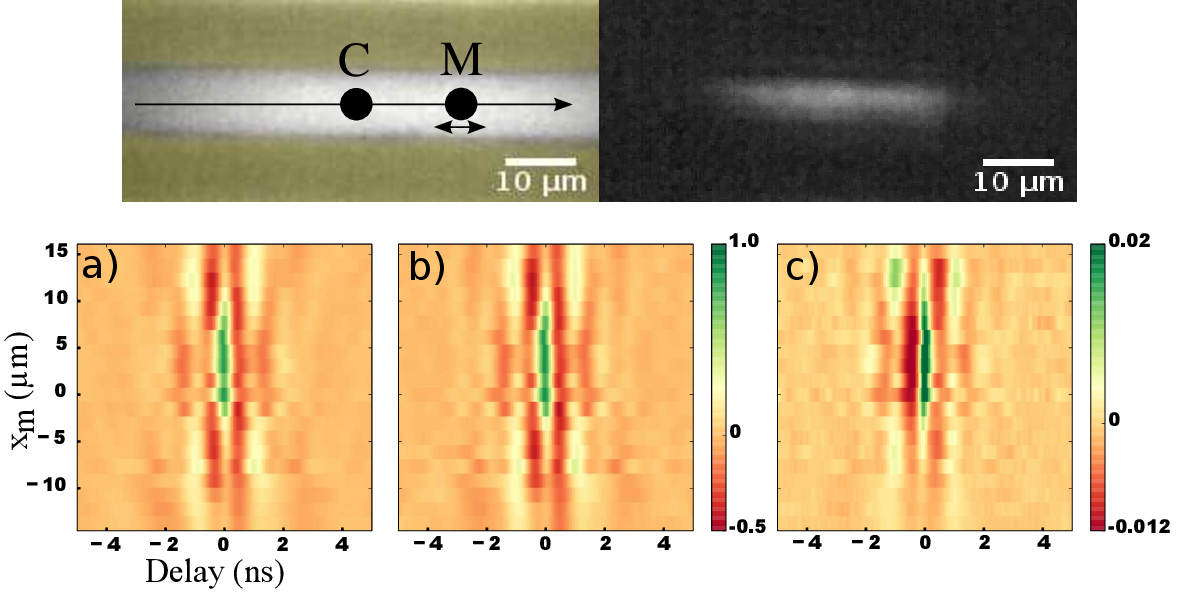}
\end{centering}
\protect\caption{\label{fig:Near-field-image}(Color online) Top panels: images of
the surface of the extended microcavity laser with integrated saturable
absorber below (left) and above (right) laser threshold. The dark
(yellow) zone is the gold mask delimiting the pumping region. Bottom
panels: a) temporal cross-correlation $X{}_{C,M}(t_{k},x_{m})$ (see
text) between the detector responses in points C ($x_{m}=0$) and
M at delays $t_{k}=k\Delta t$. b) Same as a) restricted to extreme
events at point C. c) Average of the responses at point M and at times
where an abnormal event has occurred in the center of the laser in
C. }
\end{figure}

In this Letter, we report on experimental and numerical results on
the physics of extreme events appearance in a spatially extended nonlinear
dissipative system and establish the origin of this dynamics in the
emergence of spatiotemporal chaos. Our system is a planar microcavity
laser with integrated saturable absorber \cite{BarbayAPL05,ElsassEPJD10}
pumped along a rectangular aperture, implementing a quasi 1D spatially
extended nonlinear dissipative system (cf Fig.\ref{fig:Near-field-image}).
Besides the very different dynamical regimes that can be observed
in it (e.g. laser cavity solitons \cite{ElsassAPB09,ElsassEPJD10}
or excitable regimes \cite{BarbayOL11,SelmiPRL14}), a particularity
of this system is that in absence of spatial coupling it does not
display irregular or aperiodic dynamics and hence extreme events \cite{DubbeldamOC99}.
However, spatial coupling through diffraction and nonlinear effects
can make the dynamics become more irregular, especially if the system
has a large aspect ratio (or Fresnel number) as is the case here.
Above the laser threshold, self-pulsing takes place and we study experimentally
the impact of the pumping intensity on the intensity statistics and
on the occurrence of extreme events. By recording the dynamics simultaneously
in two different spatial points we are able to study whether the extreme
events occur through a mechanism of coherent structure collision.
Indeed, stationary and propagative laser coherent structures were
predicted \cite{PerriniAPB05,RosanovAPB05,RosanovPRL05,PratiEPJD10,TissoniEPJST12,VladimirovPTRSL14}
in this system and stationary structures were observed \cite{ElsassAPB09,ElsassEPJD10}
in some parameter regions. With the help of a mathematical model,
linear stability and numerical analysis of the dynamics we unveil
the dynamical origin of the extreme events found.

The microcavity structure used in this experiment is described in
\cite{ElsassAPB09,ElsassEPJD10}. A gold mask is deposited onto the
sample surface to define the pump geometry. We concentrate on an elongated
shaped pump profile with an gold opening gold having 80$\text{\ensuremath{\mu}}$m
length and $10\mu$m width. The linear microavity is pumped above
threshold and the intensity in a point close to its center is recorded
with a fast avalanche photodiode (5GHz bandwith). The temporal signal
is amplified thanks to a low noise, high bandwidth amplifier and acquired
with a 6GHz oscilloscope at 20GS/s ($\Delta t=50$ps). Up to $50\times10^{6}$
points can be acquired in a single trace. Figure \ref{fig:Near-field-image}
shows the near field of the laser below and above threshold, respectively.

\begin{figure}
\includegraphics[width=1\columnwidth]{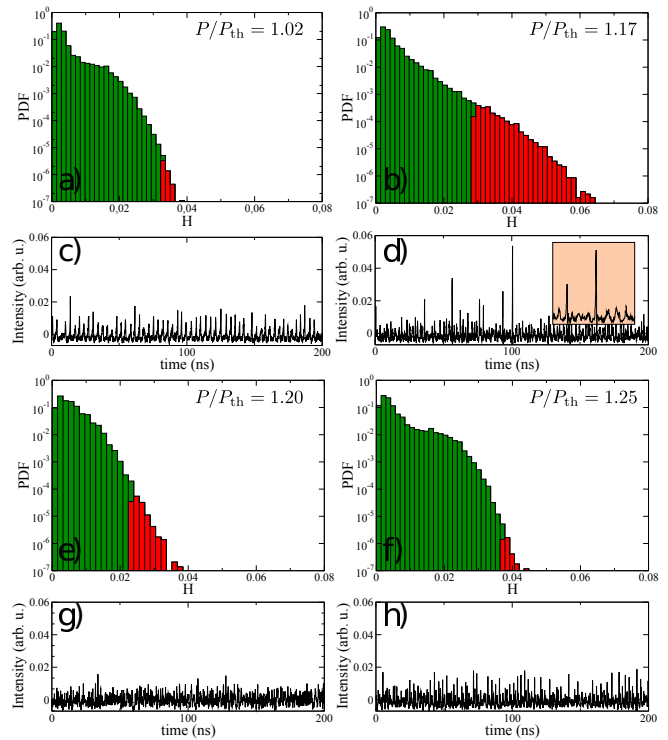}

\protect\caption{\label{fig:VsPump}(Color online) a,b,e,f) : Logarithm of the PDF
of the intensity height $H$ at position C for different normalized
pump values. Extreme events ($AI>2$) are shown in red. c,d,g,h),
are excerpts of the time evolution for the corresponding pumps. In
d) is plotted a 20ns zoom on the central extreme event. }
\end{figure}

Time traces once acquired are treated to display the histogram of
the intensity heights. Figure \ref{fig:VsPump} displays histograms
versus pump parameter. At normalized pump power $P/P_{th}=1.02$,
where $P_{th}$ is the pump at laser threshold, they are characterized
by a quadratic decay in the tails, and the probability density function
(PDF) looks like a Rayleigh distribution for a positive valued Gaussian
process. As the pump is increased, the statistics develops long tails
with an initial exponential decay ($P/P_{th}=1.17$). For still higher
pump values, the PDF becomes exponential ($P/P_{th}=1.20$) and then
redisplays Gaussian tails ($P/P_{th}=1.25$). The global evolution
of the mean amplitude versus pump intensity is reminiscent of the
dynamics expected for a zero-dimensional laser with saturable absorber \cite{TiernoPRA11}
: close to threshold, quite a regular amplitude pulse train sets in
(see Figs.\ref{fig:VsPump}c). For higher pump, the mean pulse period
increases and, because of the spatial coupling, the amplitude becomes
very irregular and displays a complex dynamics (Figs.\ref{fig:VsPump}d,g,h). We have computed
the threshold amplitude for extreme events adopting the traditional
hydrodynamical criterion. We consider as extreme events those events
having a height $H$ twice the significant height $H_{s}$ (mean of
the highest tertile of the PDF), i.e. with an abnormality index $AI\equiv H/H_{s}>2$
\cite{OnoratoPR13}. The height $H$ is extracted as the maximum of
the left and right intensity heights $H=\max(H_{l},H_{r})$. Note
that the results do not change significantly by considering either
$H$, $H_{l}$ or $H_{r}$. To get rid of the large number of small
peaks of noise at the left of the PDF, we compute the significant
height $H_{s}$ only by considering events whose height is larger
than the observed maximum peak dark noise amplitude which is about
5mV (note that the rms noise is only 0.9mV). This threshold introduces
a more stringent criterion for the extreme events detection. Extreme
events are depicted in red under the histograms presented in Fig.\ref{fig:VsPump}.
We observe, that the maximum number of extreme events is obtained
in the PDF with a non-Gaussian tail, i.e. with a normalized pump of
$1.17$.


The statistics of times between two spikes with $AI>2$ displays a
Kramers statistics with exponential behavior, marking that spikes
appearance obeys a Poisson, memoryless process. We now study the spatiotemporal
structure of the statistics of emitted pulses. We record the dynamics
in two points, one at a fixed position at the center of the laser
(represented by point C) and the other moving along the long line
laser (point M). {This is made by enlarging the laser surface image
by optical magnification and placing the detectors in that plane.}
On bottom panels in Fig.\ref{fig:Near-field-image},
we plot the normalized cross-correlation $X{}_{c,m}(k)$ of the $N=10^{5}$
first recorded points (5$\text{\ensuremath{\mu}}$s) between the signal
recorded at the central detector $y_{c}$ at point $C$ and the one
at the moving detector $y_{m}$ at location $M$, $1\le m\le20$ such
that

\[
X{}_{c,m}(k)=\frac{1}{N\sigma_{y_{c}}\sigma_{y_{m}}}\sum_{i}(y_{c}(i)-\bar{y}_{c})(y_{m}(i+k)-\bar{y}_{m})
\]
where the bar symbol and $\sigma$ indicate the mean value and the
standard deviation. In the central part appears a zone with high positive
(green) cross correlation followed and preceded by two bands of negative
cross-correlation. The temporal band in which the cross-correlation
is nonzero extends about 2ns from around zero delay. Therefore, we
can infer the existence of a finite correlation length in the system
which is smaller than the lasing system size (about 30$\text{\ensuremath{\mu}}$m).
However, since the correlation bands are vertical at these timescales,
we cannot evidence clearly propagation effects (at least with the
temporal resolution of our setup) though there is a slight bending
of the correlated band (in green). In Fig.\ref{fig:Near-field-image}b)
we restrict the cross-correlation around the points where $AI>2$,
i.e. we consider only extreme events. Notice that there are no major
differences between the two cross-correlations, hence there seems
not to be any statistical marker of the appearance of an extreme event
in this regime, and in particular no clear sign of propagation of
a coherent structure either. These results indicate that extreme height
intensity peaks appear in a spatial correlation zone and disappear
almost immediately everywhere in this zone. Correlation is therefore
maximum at zero delay for almost all positions detected. Figure \ref{fig:Near-field-image}c)
depicts the average of the responses at position $M$ and at times
where an abnormal event has occurred in the center of the laser in
$C$. The average shows a clear time asymmetry around the correlated
structure, every selected event begins with a large amplitude dip
followed by a large positive peak. On the wings of the correlated
zone we can see another dip. In this system extreme events thus appear
and disappear almost simultaneously everywhere in a correlation window.
There is no evidence, at least up to our temporal resolution, of clear
collision of coherent structures leading to the observed behavior.
Instead, we shall consider the complexity in the spatiotemporal dynamics
itself as the dynamical origin of extreme events.

\begin{figure}
\begin{centering}
\includegraphics[width=1.0\columnwidth]{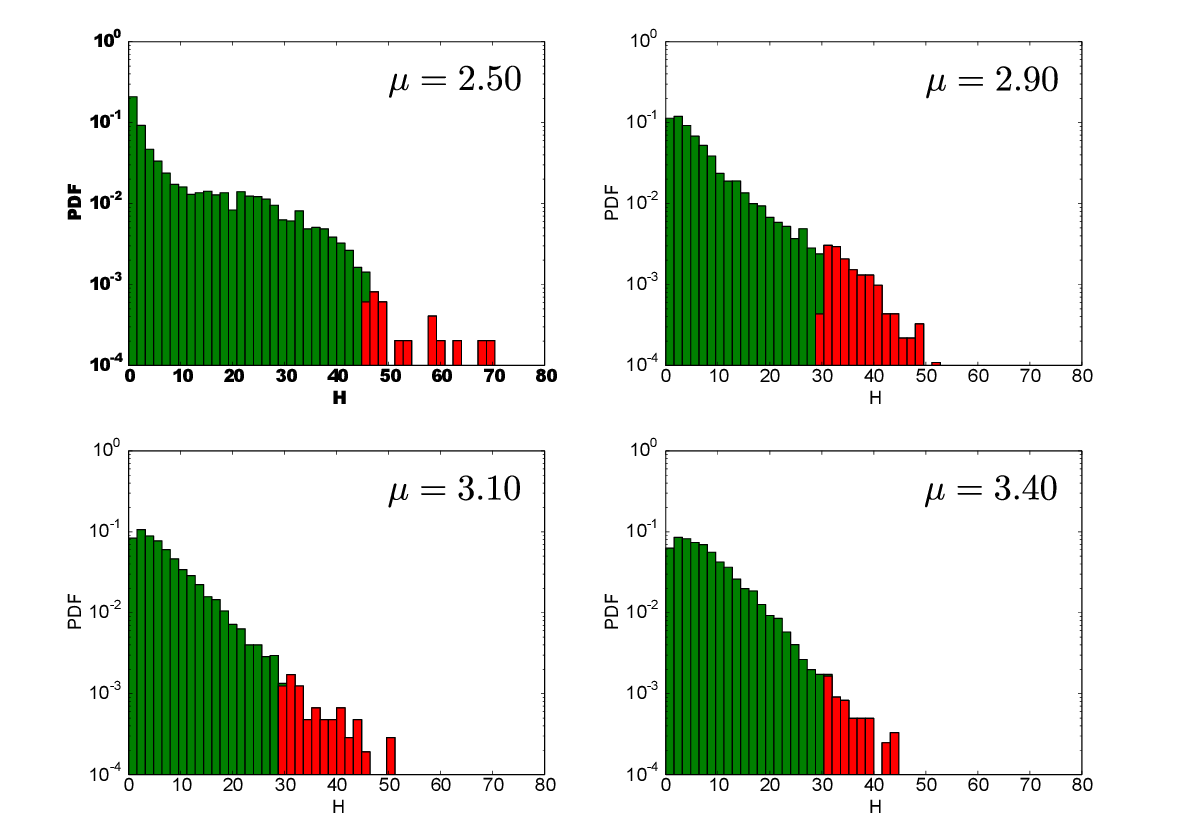}
\end{centering}

\protect\caption{\label{fig:theo-PDF}(Color online) Logarithm of the PDF of the theoretical
height distribution for the 1D laser with saturable absorber, Eqs.(\ref{eq:model}),
versus pump parameter $\mu$. Extreme events ($AI>2$) are shown in
red. }
\end{figure}

\begin{figure}
\begin{centering}
\includegraphics[width=1.0\columnwidth]{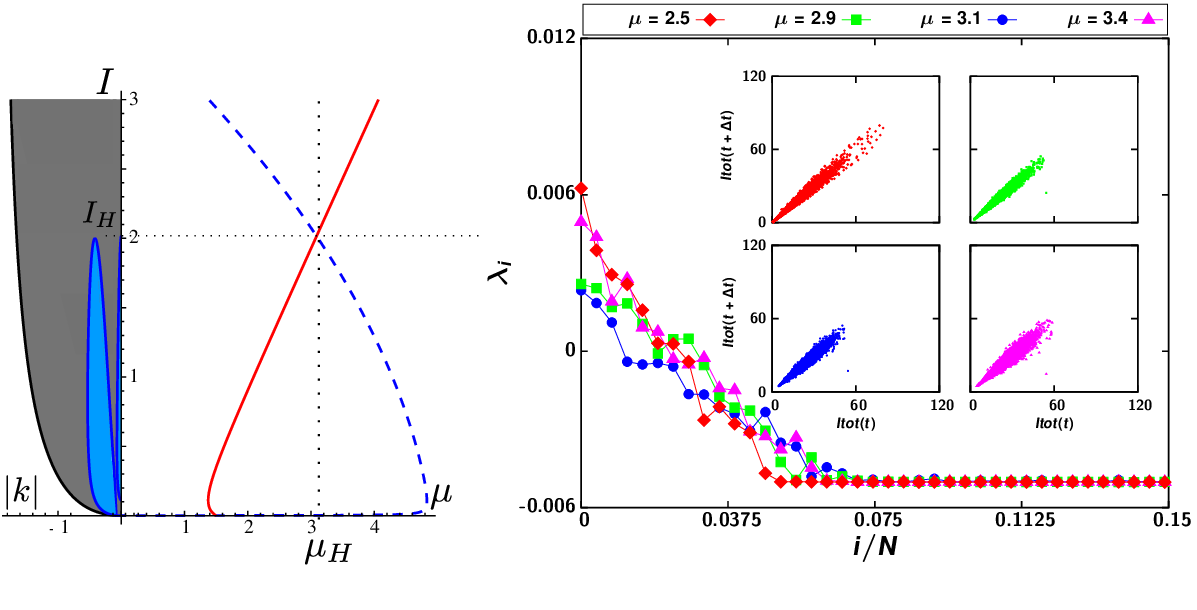}
\par\end{centering}

\protect\protect\caption{\label{fig:Turing-Lyapunov}(Color online) Phase portrait of LSA model.
Left panel stands for characteristic curve $\mu(I)$ (red) along with
the unstable wavevector regions of the the linear stability analysis
(Turing instability, grey; Andronov-Hopf instability, blue). Right
axis is $\mu$ and left axis is the modulus $\vert k\vert$ of the
unstable wavevectors. The plane-wave Hopf curve is shown in dashed,
blue. Right panel shows the computed Lyapunov spectrum for different
pump parameters and corresponding two-dimensional delay-embedding
for the total intensity $I_{tot}(t)$. }
\end{figure}

To this aim, we compare our findings with numerical simulations of
envelope equation of a one-dimensional spatially extended laser with
saturable absorber \cite{BacheAPB05}. The model consists in three
coupled nonlinear partial differential equations 
\begin{eqnarray}
\frac{\partial E}{\partial t} & = & \left[(1-i\alpha)G+(1-i\beta)Q-1\right]E+i\frac{\partial^{2}E}{\partial x^{2}}\nonumber \\
\frac{\partial G}{\partial t} & = & \gamma_{g}\left[\mu-G(1+\vert E\vert{}^{2})\right]\label{eq:model}\\
\frac{\partial Q}{\partial t} & = & \gamma_{q}\left[-\gamma-Q(1+s\vert E\vert^{2})\right]\nonumber 
\end{eqnarray}
for the intracavity electric-field envelope $E(x,t)$, the carrier
density in the gain (resp. saturable absorber) section $G(x,t)$ (resp.
$Q(x,t)$). The non-radiative carrier recombination rates are $\gamma_{g}$
and $\gamma_{q}$ with pumping $\mu$ and linear absorption $\gamma$.
The Henry enhancement factors in both sections are $\alpha$ and $\beta$,
respectively. Diffraction is included through the complex Laplacian
term. Time has been rescaled to the field lifetime in the cavity which
is calculated to be here $8.0ps$ given the cavity design parameters.
Space is rescaled to the diffraction length $w_{d}$ which is $7.4\mu m$.
We take parameters compatible with our semiconductor system : $\alpha=2$,
$\beta=0$, $s=10$, $\gamma_{g}=\gamma_{q}=0.005$ and $\gamma=0.5$.
The equations are simulated using the Xmds2 package \cite{DennisXmds2}
with a split operator method and an adaptative, fourth-order Runge-Kutta
method for time integration. The width of the integration region $w$
is $w/w_{d}=24$ with a top-hat pumping of width $w_{p}/w_{d}=12$.
Based on the results developed in \cite{BacheAPB05}, we can describe
the main properties of the plane-wave stationary solutions and of
the linear stability analysis. The results are shown on Fig. \ref{fig:Turing-Lyapunov}
for the latter set of parameters. The plane-wave characteristic curve
of the laser has a C-shape with a subcritical bifurcation at threshold
for $\mu_{th}=1+\gamma$ provided $s>1+1/\gamma$. In a certain range of
parameters, the system also exhibits an Andronov-Hopf bifurcation
giving rise to self-pulsation (for $\mu<\mu_{H}\sim3.08$). When including
the spatial degree of freedom, a linear stability analysis reveals
that the upper branch is usually Turing unstable everywhere (gray
region), giving rise to a complex spatiotemporal dynamics. A Andronov-Hopf
instability can also occur for small harmonic perturbations in space
with a band of unstable wavevectors $k$ (blue region disconnected
from the vertical axis).

The Logarithm of the PDF for the theoretical height distribution for
Eqs.(\ref{eq:model}) is shown in Fig.\ref{fig:theo-PDF}. For low
pumping it displays a sub-exponential tail with a small number of
extreme events. Then the tail of the PDF progressively becomes more
and more exponential at the start of the distribution with a large
deviation for large events giving rise to a maximum number of extreme
events for $\mu=2.9$. The tail of the distribution becomes then quasi
exponential at $\mu=3.1$ and then sub-exponential again at $\mu=3.4$
with a decrease of the number of extreme events. These observations
reproduce qualitatively well what is found in the experiment. Moreover,
the shape of the distribution seems to be strongly correlated to the
presence or not of a Andronov-Hopf bifurcation : only when it is present
can we observe a heavy tailed distribution. At the transition between
the Hopf-Turing and Turing-only region we observe the maximum number
of extreme events (for $\mu=2.9$).

A characterization of chaos and spatiotemporal chaos can be achieved
by means of Lyapunov exponents \cite{MannevilleBook90}. These exponents
measure the growth rate of generic small perturbations around of a
given trajectory in a finite dimensional dynamical systems. There
are as many exponents as the dimension of the system under study.
Additional information about the complexity of the system can be obtained
from the exponents, for instance the dimension of the strange attractor
(spectral dimensionality) or measures of the dynamic disorder (entropy)\cite{OttBook02}
or characterization of bifurcations diagram \cite{ClercPRE03}. The
analytical study of Lyapunov exponents is a thorny endeavor and in
practice inaccessible. Hence, a reasonable strategy is to derive the
exponents numerically by discretizing the set of partial differential
equations (\ref{eq:model}). Let $N$ be the number of discretization
points, then the system has $N$ Lyapunov exponents $\lambda_{i}$.
If the Lyapunov exponents are sorted in decreasing order and in the
thermodynamic limit ($N\to\infty$), these exponents converge to a
continuous spectrum as Ruelle conjectured \cite{RuelleCMP82}. Therefore,
if the system has spatiotemporal chaos in this limit, there exists
an infinite number of positive Lyapunov exponents. The set of Lyapunov
exponents provides an upper limit for the strange attractor dimension
through the Kaplan-Yorke dimension \cite{OttBook02} $D_{\mathrm{KY}}=p+\sum_{i-1}^{p}\lambda_{i}/\lambda_{p+1}$,
where $p$ is the largest integer that satisfies $\sum_{i-1}^{p}\lambda_{i}>0$
. In the thermodynamic limit the Yorke-Kaplan dimension diverges with
the size of the system as a consequence of the Lyapunov density \cite{PaulPRE07}.
We have calculated the Lyapunov spectrum (cf. Fig.\ref{fig:Turing-Lyapunov})
corresponding to the total intensity integrated over x in the model
(\ref{eq:model}). This figure clearly shows that when the system
exhibits extreme events it is in a regime of spatiotemporal chaos
with several non-zero Lyapunov exponents in the Lyapunov spectrum
and an absence of structure in the delay embedding.

\begin{figure}
\begin{centering}
\includegraphics[width=1.0\columnwidth]{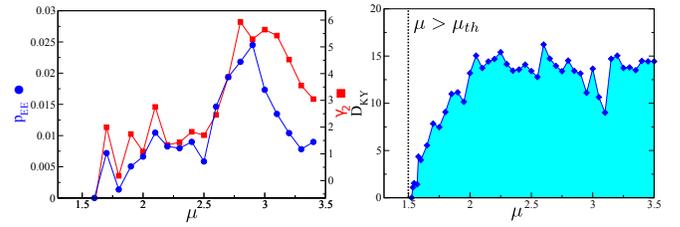} 
\end{centering}
\protect\caption{\label{fig:statistical-indicators}(Color online) Left panel : Proportion
of extreme events ($p_{EE}$, blue circles), normed kurtosis ($\gamma_{2}$,
red squares) versus pump $\mu$. Right panel : Kaplan-Yorke dimension
($D_{\mathrm{KY}}$, blue diamonds) versus pump $\mu$. }
\end{figure}

Moreover, we have computed the proportion of extreme events $p_{EE}$,
the normed kurtosis $\gamma_{2}=\mathbb{E}\left[\left(\left(X-\mu\right)/\sigma\right)^{4}\right]-3$
and the Kaplan-Yorke dimension $D_{\mathrm{KY}}$ versus pump in Fig.\ref{fig:statistical-indicators}.
{$p_{EE}$ and $\gamma_2$} both display a maximum versus pump around $\mu\simeq3$
with some correlated oscillations. {$D_{\mathrm{KY}}$} increases steadily from
zero at $\mu=1.525$ and then saturates after $\mu=2$. From these
findings we infer that there is a smooth or supercritical transition
of the system into spatiotemporal chaos and this behavior is concomitant
with the increase of the number of extreme events. Note however that
there is no reason why there should be a strict correlation between
$D_{KY}$ and $p_{EE}$ since the latter is related to the structure
of the attractor itself and not only to its dimension {\cite{LucariniJSP14}}. 

In conclusion, we have shown experimental results of extreme events
appearance in a quasi-1D broad area laser with saturable absorber.
We have analyzed the physical
origin of extreme events that occur because of the onset of {deterministic} spatiotemporal
chaos in the system. 
{Irregular dynamics is obviously a prerequisite for the observation of extreme events but we show in our work that the proportion of extreme events is not directly linked to the evolution of the Kaplan-Yorke dimension. A higher dimensional dynamics does not lead necessarily to a higher number of extreme events.}
The origin of extreme events in that case is thus to be found in the nature of
the spatiotemporal complexity that takes place, and thus could offer
interesting prospects for control through changing the system geometry
or the nature of the coupling.

\section*{Acknowlegements}

F.S., S.C., Z.L. and S.B. acknowledge partial support from the French
network Renatech and the ANR Blanc project Optiroc. M.G.C. thanks
for the financial support of FONDECYT projects 1150507.


%

\end{document}